\documentclass[RNAAS]{aastex62}
\usepackage{CJKutf8}

\graphicspath{{./}{figures/}}

\begin{document}

\title{The Distance to NGC\,1042 in the Context of its Proposed Association with the Dark Matter-Deficient Galaxies NGC1052-DF2 and NGC1052-DF4}

\correspondingauthor{Pieter van Dokkum}
\email{pieter.vandokkum@yale.edu}

\author{Pieter van Dokkum}
\affiliation{Yale University}


\author{Shany Danieli}
\affiliation{Yale University}

\author{Aaron Romanowsky}
\affiliation{San Jos\'e State University}
\affiliation{University of California Observatories}

\author{Roberto Abraham}
\affiliation{University of Toronto}


\author{Charlie Conroy}
\affiliation{Harvard-Smithsonian Center for Astrophysics}







\gdef\gal{NGC\,1042}
\gdef\kms{km\,s$^{-1}$}
\gdef\msun{M$_{\odot}$}
\section{A ``short distance" solution to NGC1052-DF2 and NGC1052-DF4} 

The 
galaxies NGC1052-DF2 \citep{vd18a} and NGC1052-DF4
\citep{vd19} have received considerable attention, as their kinematics
indicate that they have little or
no dark matter.
Both galaxies are also unusually large for their mass,
and both host a population of unusually luminous
globular clusters.

Most of the properties of the
galaxies are distance-dependent. The distances were
measured to be $19-20$\,Mpc
\citep{cohen:18}, consistent with
that of the nearby elliptical galaxy NGC\,1052 and with their
radial velocities ($cz=1804$\,\kms\ for NGC1052-DF2 and $cz=1445$\,\kms\ for
NGC1052-DF4).  However, \citet{trujillo:19} suggest a shorter distance of
$\approx 13$\,Mpc: for that
distance the globular clusters and velocity dispersions are
still highly unusual, but less extreme than for 19--20\,Mpc.


In \citet{vd18b} we analyzed this
suggestion,\footnote{The Trujillo et al.\ study is still under review, but appeared on the arXiv in June 2018.}
showing that the CMD
of NGC1052-DF2 is inconsistent with a distance of 13\,Mpc and
that its surface brightness fluctuation signal implies
$D=18.7\pm 1.7$\,Mpc independent of absolute
calibrations. Here we focus on
a different aspect of the \citet{trujillo:19} argument.
NGC1052-DF2 and NGC1052-DF4 are likely at the same distance, but their
velocity difference of 360\,\kms\ rules out the
possibility that they are an isolated bound pair.
Therefore, the short distance solution
only ``works" if they are satellites
of a massive galaxy in the
foreground of the NGC\,1052 group.
The only plausible candidate is the spiral galaxy NGC\,1042; in projection
it is near
NGC1052-DF2 and NGC1052-DF4, and Trujillo et al.\ suggest its distance is
``within a range of 8--13\,Mpc". 

\begin{figure}[htbp]
\begin{center}
\includegraphics[scale=0.5,angle=0]{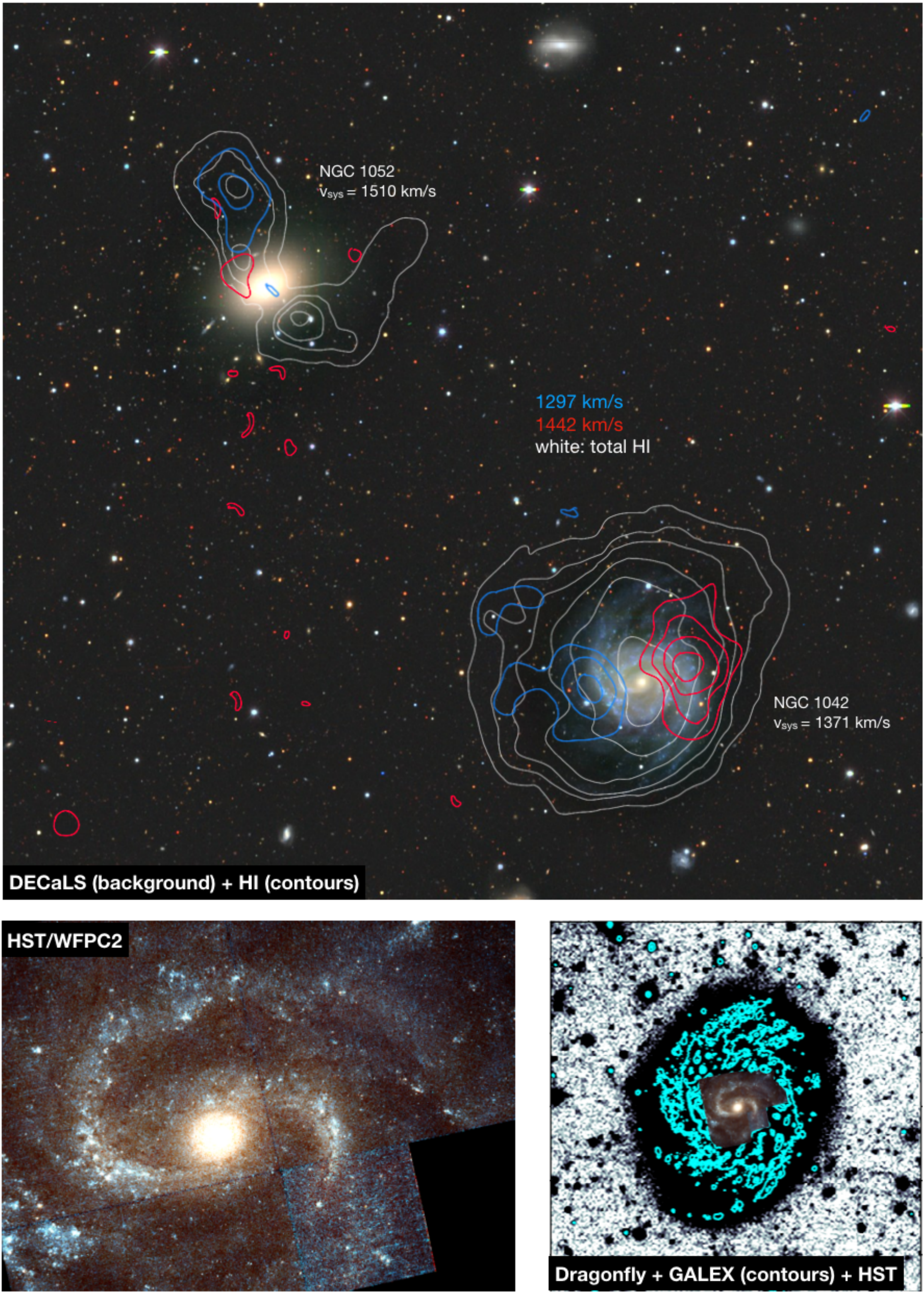}
\caption{{\em Main panel:} H\,{\sc i} distribution 
adapted from \citet{vgork}.
{\em Lower left:}
HST/WFPC2 image of the central regions of
NGC\,1042. 
{\em Lower right:} Large
scale distribution of star formation (blue)
and stars (grey). 
}
\end{center}
\end{figure}

\section{Neutral hydrogen
in NGC\,1042 and NGC\,1052}

In this {\em Research Note}
we show that NGC\,1042 is probably not in the foreground
of the NGC\,1052 group but a member of it. In projection it is near
NGC\,1052 itself, its radial velocity of
$cz=1371$\,\kms\ is close to the central group
velocity ($cz\approx 1450$\,\kms), and in group catalogs such as that of \citet{kourkchi:17} it is listed as a bona fide member. An
additional argument comes from
the H\,{\sc i} distribution in NGC\,1042 and NGC\,1052, as shown in \citet{vgork}. NGC\,1052 has H\,{\sc i} emission in two spatially-
and dynamically-distinct
components, to the NE and SW of the galaxy. The velocity
range of the NE component  (1297\,\kms\,--\,1442\,\kms)
overlaps precisely with that of NGC\,1042
\citep[see Figs.\ 1 and 15 of][]{vgork}. Furthermore, the
H\,{\sc i} in both galaxies has a disturbed morphology. We show
the total H\,{\sc i} emission in Fig.\ 1,
as well as channel maps at 1297\,\kms\ and 1442\,\kms.
The blue- and redshifted velocities of NGC\,1042 show ordered rotation
but the total
emission exhibits an asymmetry along the NE-SW axis, in the direction of NGC\,1052.
As noted
in \citet{vgork} NGC\,1042 is the only galaxy in the group with a large
gas reservoir, and
a likely interpretation of Fig.\ 1 is that NGC\,1052 stripped
gas from NGC\,1042 in a past interaction. 

\section{The Tully-Fisher Distance to NGC\,1042}

The distance of NGC\,1042 quoted in Trujillo et al.\ (``8--13\,Mpc") is based on the Tully-Fisher
relation. The NASA/IPAC Extragalactic Database lists 17 Tully-Fisher distances for NGC\,1042,
ranging from 4.2\,Mpc to 16.7\,Mpc. The main reason for this large range is the
uncertain inclination of the galaxy.
With a velocity width of $W_{mx}=V_{\rm{}max}-V_{\rm min}\approx{}145(\sin i)^{-1}$\,\kms\ \citep{vgork,luo:16},
an observed magnitude of $m_I^{\rm{}T}=10.19$, and the RC3 inclination
of $i=40\arcdeg$, the \citet{tf} calibration gives
$D_{\rm{}TF}\approx11.5$\,Mpc.
The RC3 inclination is derived from the
axis ratio, but as shown in the bottom panels of Fig.\ 1 the
axis ratio and position angle of NGC\,1042 are
strongly dependent on radius. In the inner regions the axis
ratio is strongly influenced by the bar. At larger radii the
galaxy first becomes nearly round and then flattens again but now {\em in the opposite direction}, such that it is elongated
along the rotation axis in the Dragonfly data. These results reinforce
the notion that the galaxy is disturbed, and also suggest that it
is viewed close to face-on. For inclinations $i<30\arcdeg$ the Tully-Fisher
distance to NGC\,1042 is $D_{\rm{}TF}\gtrsim{}18$\,Mpc. 

\section{Conclusion}

We infer that
NGC\,1042 is almost certainly a member of the
NGC\,1052 group, consistent with previous determinations
that were based on its radial velocity and location in the sky \citep{kourkchi:17}.




\acknowledgments
We thank Jacqueline van Gorkom for comments on the manuscript.

\end{document}